\documentclass[sigconf,screen]{acmart}

\makeatletter                  
\def\mdseries@tt{m}      
\makeatother                   

\usepackage{hyperref} 
\usepackage{amssymb} 
\usepackage{xspace}
\usepackage{color}
\usepackage{xcolor}
\usepackage{fancybox}
\usepackage{fancyhdr}
\usepackage{ifthen} 
\usepackage{paralist} 

\usepackage[T1]{fontenc}
\usepackage{multirow}

\usepackage{algorithm}
\usepackage[noend]{algpseudocode}
\algnewcommand{\LineComment}[1]{\State \(\triangleright\) #1}

\usepackage{amsmath} 

\usepackage{flushend} 
\newboolean{acmtemplate}
\setboolean{acmtemplate}{true} 
\newboolean{anonymous}
\setboolean{anonymous}{false}
\newboolean{includecomments}
\setboolean{includecomments}{true}

{{\ttfamily \hyphenchar\the\font=`\-}

\newcommand{\paperTitle}{Redefine Paper Title Later}
\newcommand{\submitTo}{ACRONYM of conference later}
\newcommand{\maxPagesAllowed}{?? to be redefined}

\clubpenalty = 10000
\widowpenalty = 10000
\displaywidowpenalty = 10000

\hypersetup{
    colorlinks,%
    citecolor=black,%
    filecolor=black,%
    linkcolor=black,%
    urlcolor=gray,
    linktocpage=true,
    bookmarks=true,
    bookmarksopen=true,
    pdfpagemode=UseOutlines,
    pdftitle={\paperTitle},    
    pdfauthor={Anonymous},     
    pdfsubject={Draft for \submitTo Conference},   
    pdfcreator={PDF LaTex},   
}

\setlength{\textfloatsep}{2mm} 
\newcommand{\hypobox}[1]{
	\begin{center}%
        \noindent\thicklines\setlength{\fboxsep}{2pt}%
        \cornersize{0.1}
        \ovalbox{\begin{minipage}{8.3cm}%
		#1
		\end{minipage}}
	\end{center}}

\ifthenelse{\boolean{anonymous}} {

}{

}

\newcommand{\secref}[1]{Section~\ref{#1}\xspace}
 
\newcommand{\tabref}[1]{Table~\ref{#1}\xspace}
\newcommand{\algoref}[1]{Algorithm~\ref{#1}\xspace}

\ifthenelse{\boolean{includecomments}} {
    \newcommand{\nb}[3]{
    	{\colorbox{#2}{\bfseries\sffamily\scriptsize\textcolor{white}{#1}}}
    	{\textcolor{#2}{\sf\small$\blacktriangleright$\textit{#3}$\blacktriangleleft$}}}
} {
    \newcommand{\nb}[3]{}
}

\newcommand*{\RQOne} [1] {Isn't this a splendid research question here ?}
\newcommand*{\RQTwo} [1] {And isn't this one even better ?}
\newcommand*{\RQThree} [1] {This one tops it all, doesn't it ?}

\setcopyright{none}

\begin{document}
\renewcommand{\paperTitle}{Goal-Oriented Mutation Testing with Focal Methods}
\renewcommand{\submitTo}{A-TEST 2018}
\renewcommand{\maxPagesAllowed}{7}
\title{\paperTitle} 

    {
        \author{Sten Vercammen}
        \affiliation{%
        \institution{University of Antwerp}
        \streetaddress{Middelheimlaan 1}
        \city{Antwerp} 
        \country{Belgium}
        \postcode{2020}
        }
                
        \author{Mohammad Ghafari}
        \affiliation{%
        \institution{University of Bern} 
        \streetaddress{Hochschulstrasse 6}
        \city{Bern} 
        \country{Switzerland}
        \postcode{CH-3012}
        }

        \author{Serge Demeyer}
        \orcid{0000-0002-4463-2945}
        \affiliation{%
        \institution{University of Antwerp}
        \streetaddress{Middelheimlaan 1}
        \city{Antwerp} 
        \country{Belgium}
        \postcode{2020}
        }

     	\author{Markus Borg}
        \affiliation{
        \institution{RISE SICS AB}
        \streetaddress{Scheelev\"{a}gen 17}
        \city{Lund} 
        \country{Sweden} 
        \postcode{SE-223 70}
        }
                
        \renewcommand{\shortauthors}{Vercammen et al.}
    }

\begin{abstract}
Mutation testing is the state-of-the-art technique for assessing the fault-detection capacity of a test suite. Unfortunately, mutation testing consumes enormous computing resources because it runs the whole test suite for each and every injected mutant. In this paper we explore fine-grained traceability links at method level (named focal methods), to reduce the execution time of mutation testing and to verify the quality of the test cases for each individual method, instead of the usually verified overall test suite quality. Validation of our approach on the open source Apache Ant project shows a speed-up of 573.5x for the mutants located in focal methods with a quality score of 80
\end{abstract}


\ifthenelse{\boolean{acmtemplate}} {
\maketitle
}
{
\IEEEpeerreviewmaketitle
\maketitle
}


\section{Introduction}
\label{sec::Intro}
Software testing is the dominant method for quality assurance and control in software engineering~\cite{ng2004preliminary,garousi2013survey}, established as a disciplined approach already in the late 1970's. Originally, software testing was defined as \emph{``executing a program with the intent of finding an error''}~\cite{myers197977ie}. In the last decade, however, the objective of software testing has shifted considerably with the advent of continuous integration~\cite{adams2016modern}. Many software test cases are now fully automated, and serve as quality gates to safeguard against programming faults. 

Large-scale test automation is now a common practice among mature software-intensive businesses. For example, Microsoft reported that approximately 11 months of development on Windows comprised more than 30 million test case executions and Google stated that \emph{``In an average day, TAP integrates and tests [\ldots] more than 13K code projects, requiring 800K builds and 150 Million test runs.''}~\cite{Memon2017googlescale}. By adopting high quality software testing in the continuous integration context, software companies are now releasing software much more frequently. Examples include Tesla, uploading new software in their cars once every month~\cite{TeslaUpdates}, and Amazon, pushing new updates to production every 11.6 seconds~\cite{Jenkins2011}.

Test automation is a growing phenomenon in industry, but a fundamental question remains: How trustworthy are these automated test cases? Mutation testing is currently the state-of-the-art technique for assessing the \textit{fault-detection capacity} of a test suite~\cite{jia2011analysis}. The technique systematically injects faults into the system under test and analyses how many of them are killed by the test suite. In the academic community, mutation testing is acknowledged as the most promising technique for automated assessment of the strength of a test suite~\cite{Papadakis2019}. One of the major impediments to industrial adoption of mutation testing is the computational costs involved: each individual mutant must be deployed and tested separately~\cite{jia2011analysis}.

For the greatest chance of detecting a mutant, the entire test suite is executed for each and every mutant~\cite{chekam2017empirical}. As this consumes enormous resources, several techniques to exclude test cases from the test suite for an individual mutant have been proposed.
First and foremost, test prioritisation reorders the tests cases to first execute the test with the highest chance to kill the mutants~\cite{zhang2013faster}. Second, program verification excludes test cases which cannot reach the mutant and/or which cannot infect the program state~\cite{bardin2015sound}. 
Third, (static) symbolic execution techniques identify whether a test case is capable of killing the mutant~\cite{papadakis2012mutation, holling2016nequivack}.
This paper explores an alternative technique: fine-grained traceability links via \textit{focal methods}~\cite{ghafari2015automatically}.

By using focal methods, we can establish a traceability link at method level between production code and test code. This allows us to identify which test cases actually test which methods and vice versa. The greatest advantage of this technique is that if we know which test cases focus on testing which methods, we do not need to execute the whole test suite nor test cases that only cover a method.
This allows us to drastically reduce the scope of the mutation analysis to a fraction of the entire test suite by executing only those test cases that actually test the methods of interest. This technique can also be used to quickly investigate how well a single method is tested by only executing the mutants of that method with the (few) test cases that actually test the method.
We refer to this as goal-oriented mutation testing, and argue that the approach could be used to selectively target the parts of a system where mutation testing would have the largest return on investment.

\vspace{0.5em}
\noindent
To investigate the potential of focal methods in the context of mutation testing, we formulate the following research questions.

\begin{compactitem}
\item \textbf{RQ1:} \textit{To what extent, using focal methods, can we identify the right mutants for a test case and vice versa?}

\item \textbf{RQ2:} \textit{How much speed-up is gained by using focal methods for mutation testing?}

\end{compactitem}

We validate this concept on the unit testing level using a large open source project: Apache Ant [\url{https://ant.apache.org}].

The rest of the paper is structured as follows.
In \secref{sec:Background}, we elaborate on the concept of mutation testing and general optimisations. In \secref{sec:Goal-Oriented Mutation Testing}, we describe the motivation for using focal methods, and how they work.
In \secref{sec:Case Study}, we explain our case study setup and discuss the results.
In \secref{sec::RelatedWork}, we elaborate on related work.
As with any empirical research, our study is subject to threats to validity -- we list the most important in \secref{sec::Threats}. Finally, we arrive at a conclusion in \secref{sec::Conclusion}.


\section{Background}
\label{sec:Background}
This section explains why we need mutation testing, why it needs to be optimised, and which types of optimisations exist.

\subsection{Mutation Testing}
Software teams need effective test cases to maximise the likelihood of exposing faults~\cite{myers197977ie}. Traditionally, code coverage has been used to assess the strength of a test suite, revealing which parts of the production code are inadequately tested. Unfortunately, research has shown that code coverage might be a poor indicator of test effectiveness~\cite{cai2005effect, inozemtseva2014coverage}. On top of that, even a 100\% MC/DC coverage (Modified Condition/Decision Coverage, the coverage criterion mandated by safety standards used for certification of safety-critical systems) does not guarantee the absence of faults~\cite{Gay2015riskcoverage, Kandl2015}.

Mutation testing is today the state-of-the-art technique for assessing the \textit{fault-detection capacity} of a test suite~\cite{jia2011analysis, Papadakis2019}. By deliberately injecting faults (called \textit{mutants}) into the production code, and counting how many of them are killed by the test suite, mutation testing has been shown to outperform traditional code coverage approaches. Case studies with safety-critical systems demonstrate that mutation testing could be effective where traditional code coverage analysis and manual inspections fail~\cite{Baker2013mutationtestingsafetrycritical, Ramler2017mutationtestingsafetrycritical}. Furthermore, Google reports that mutation testing both can provide insight into poorly tested parts of the system, and also reveal design problems with modules that are difficult to test, i.e. mutation testing can identify candidates for refactoring~\cite{Petrovic2018mutationtestingatgoogle}.

Still, mutation testing is rarely adopted in industry practice~\cite{gopinath2014code}. One of the reasons is that mutation testing traditionally is computationally expensive, as the code base must be compiled and tested separately for each mutant~\cite{jia2011analysis}.
\algoref{code:Mutation Testing} shows the fundamental steps of mutation testing.
As a prerequisite for mutation testing, referred to as the \textit{pre}-phase, the software system needs to build without errors, and all software test cases should execute successfully. Subsequently, the two main phases are executed: (\textit{A}) the mutant generation phase and (\textit{B}) the mutant execution phase. In phase \textit{A}, mutants are generated for all source code files. In phase \textit{B}, test cases for each mutant are executed and the result (whether or not the mutant was killed) is saved. As a final step, the \textit{post}-phase, all results are collected and the final report is created.

\begin{algorithm}[hbt]
\caption{Pseudocode Mutation Testing}
\label{code:Mutation Testing}
\begin{algorithmic}[1]
\Function{mutationTesting}{srcFolder $src$}
	\LineComment{Pre: verify build and if all tests succeed}
	\If {$\Call{initialBuildAndTests}{ } \not= \textbf{True}$}
		\State \Return
	\EndIf
	\State
	\LineComment{A: generate mutants}
	\State $mutants\gets []$
	\ForAll{srcFile $f \in src$}
		\State $fMutants\gets \Call{generateMutants}{$srcFile $f}$
		\State $mutants\gets mutants + fMutants$
	\EndFor
	\State
	\LineComment{B: execute mutants}
	\ForAll{mutant $m \in mutants$}
		\State $result\gets \Call{executeMutant}{$mutant $m}$
		\State \Call{storeResult}{$result$, mutant $m$}
	\EndFor
	\State
	\LineComment{Post: process results}
	\State $\Call{processResults}{ }$
\EndFunction
\end{algorithmic}
\end{algorithm}

\subsection{Mutation Testing Optimisations}

A lot of research is devoted to optimising the mutation testing process, summarised under the vision - \textit{do fewer}, \textit{do smarter}, and \textit{do faster}~\cite{offutt2001mutation}. 

The \textit{do fewer} approaches minimise the execution time by reducing the total amount of mutants to execute. Such an optimisation can be implemented by generating fewer mutants on line 9 in \algoref{code:Mutation Testing} or by selecting a subset of all mutants on line 13. The fewer mutants that are executed, the more information will be lost. Balancing time reduction versus information loss is key. There are different ways to choose which mutants will be executed, varying in their effectiveness compared to the full set of mutants~\cite{jia2011analysis}.

\textit{Do smarter} approaches attempt to minimise the execution time by retaining state information between runs, e.g. split-stream mutation testing~\cite{king1991fortran}. Another example is \textit{test prioritisation}, which gives priority to the test cases with the highest likelihood of failure. These optimisation would be implemented on line 14 in \algoref{code:Mutation Testing}. 

Lastly, \textit{do faster} approaches try to minimise the execution time of each individual mutant. One example is using a compiler integrated technique, where the project is compiled only once instead of for each mutant~\cite{demillo1991compiler}. These optimisation would also be implemented on line 14 in \algoref{code:Mutation Testing}.

Currently, optimisation techniques with large speedups sacrifice accuracy~\cite{jia2011analysis, Papadakis2019}. Thus when evaluating mutation optimisation techniques, the trade-off between speed-up and accuracy must be quantified.


\section{Goal-oriented Mutation Testing}
\label{sec:Goal-Oriented Mutation Testing}
Mutation testing and mutation testing optimisations have focused on detecting the \emph{overall} quality of the test suite as fast as possible. We, however, propose a more focused approach to mutation testing, where only the test cases that actually test a method are considered.

\subsection{Motivation}
\label{subsec:Motivation}
Finding the root cause of a fault is not an easy task, an entire field of study is dedicated specifically to this problem. One solution is spectrum-based fault localisation. It tries to locate the faulty component by cross referencing the test cases which detect the fault and which components are used in those test cases. One of the most recent advances uses a new metric called DDU (Density-Diversity-Uniqueness) to assess the diagnostic ability of a test suite~\cite{perez2017test}. 
The authors state that \textit{``[t]he metric, tries to emulate the properties of calculating per-test coverage entropy, to ensure accurate diagnosability. Ideal diagnostic ability can be proved to exist when a suite reaches maximum entropy
...
DDU focuses on three particular properties of entropy by ensuring that a) test cases are diverse; b) there are no ambiguous components; c) there is a proportional number of test cases of distinct granularity; while still ensuring tractability.''
}
They observed a statistically significant increase in diagnostic performance of about 34\% when locating faults by optimising DDU compared to branch-coverage. 

In essence, this avoids the \textit{eager test} code smell~\cite{van2001refactoring}, i.e. testing too many methods of an object in a single test case. To increase the diagnosability of the test cases, this and other forms of \textit{test smells} should be avoided~\cite{van2001refactoring}. From this we can conclude that it is best to test a method $f$ in a test case that is specifically designed to test $f$ and not in a test case that just so happens to call the method $f$ in one of its routines.

If method $f$ would be faulty, then $testF$ is responsible to detect this. If $f$ calls methods $a, b$, and $c$, then $testA, testB$, and $testC$ are responsible for detecting faults in their respective methods. Therefore, $testF$ is not required to detect a fault if the fault is inside method $a, b$, or $c$. We thus argue that given a faulty method $f$, it should suffice to only execute those test cases which are responsible for testing the method $f$.
Finding which test cases are responsible to test a method can be achieved using \textit{focal methods}~\cite{ghafari2015automatically}.

\subsection{Focal Methods}
Method invocations within a test case play different roles in the test.
A majority of them are ancillary to a few ones that are intended to be the actual (or focal) methods under test.
More particularly, unit test cases are commonly structured in three logical parts: setup, execution, and oracle.
The setup part 
instantiates the class under test, and includes any dependencies on other objects that the unit under test will use. This part contains initial method invocations that bring the object under test into a state required for testing. The execution part
stimulates the object under test via a method invocation, i.e., the \emph{focal method}~\cite{ghafari2015automatically, ghafari2017mining} in the test case. This action is then checked with a series of inspector and assert statements in the oracle part
that controls the side-effects of the focal invocation to determine whether the expected outcome is obtained.


\algoref{code:Test Savings Example} represents a unit test case of a savings account where the intent is to test the \textit{withdraw} method. For this, an account to test the \textit{withdraw} method must exist. Thus, an account is created on line 3 (in \algoref{code:Test Savings Example}). To deposit or withdraw money to/from an account, the user must first authenticate himself (line 4). To make sure that the account has money to withdraw, a deposit must be made (line 5). If the savings account has a sufficient amount of money, the \textit{withdraw} method will withdraw the money from the account (line 7), and the remaining amount of money in the savings account should be reduced. The latter is verified using the \textit{assert} statement on line 11.

In the example, the intent clearly is to test the \textit{withdraw} method. 
This method is the focal method as it is the last method that updates the internal state of the $account$ object. The expected change is then evaluated in the oracle part by observing the result of the focal method (line 10), as well with the help of the $getBalance$ method which only inspects the current balance.  
\renewcommand*\Call[2]{\textproc{#1}(#2)}
\begin{algorithm}[hbt]
\caption{Exemplary Unit Test Case for Money Withdrawal}
\label{code:Test Savings Example}
\begin{algorithmic}[1]
\Function{testWithdraw}{}
	\LineComment{Setup: setup environment for testing}
	\State $account\gets$\Call{createAccount}{$account$, $auth$}
	\State $account$.\Call{authenticate}{$auth$}
	\State $account$.\Call{deposit}{$10$}
	\LineComment{Execution: execute the focal method}
	\State $success\gets account$.\Call{withdraw}{$6$}
	\LineComment{Oracle: verify results of the method}
	\State $balance\gets account$.\Call{getBalance}{}
	\State \Call{assertTrue}{$success$}
	\State \Call{assertEqual}{$balance$, $4$}
\EndFunction
\end{algorithmic}
\end{algorithm}

Therefore, focal methods represent the core of a test scenario inside a unit test case. Their main purpose is to affect an object's state that is then checked by other inspector methods whose purpose is ancillary.
A tool to detect focal methods exists, and has been recently used for extracting API usage examples from unit test cases~\cite{ghafari2017mining}.

\subsection{Limiting the Test Scope}
\label{subsubsec:limiting}
Under the premise that it is the responsibility of the (few) test cases that test a focal method $f$ to catch all faults in the method $f$, we can assume that it suffices to limit the test scope to these selected test cases when we are looking for faults in method $f$. We assume even if we exclude those test cases that only happen to call a method $f$ in one of its routines, there ought to be a simpler test case which tests the method $f$ as a focal method that ought to also reveal that the method is faulty as that test case bears the responsibility to test the method and not the more complex test case.

Applied to mutation testing, this means that if a mutant is located in method $f$, we only need to execute these (few) test cases that test the focal method $f$.

It is possible that some test cases that are designed to test a particular method do not kill a mutant while the complete test suite can. We can define the \textit{quality} of the individual test cases by investigating how many of the mutants that are located in a method are killed by the test cases that are designed to test the particular method. This quality score can be calculated by dividing the amount of mutants killed by this technique with the total amount of mutants located in the focal methods.


\hypobox{\textbf{Summary.} We propose a more straightforward approach to mutation optimisation that incorporates the diagnostic capabilities of test suites. By only executing the test cases which actually are meant to test the method $f$, i.e. the focal method, we hope to drastically reduce the amount of test cases needed for mutants located in method $f$, reducing the execution time, while retaining the fault detection capabilities of the test suite.}

\section{Case Study}
\label{sec:Case Study}
In this section, we explain our case setup, why and how we want to investigate our research questions, explain our gathered results, and answer our research questions accordingly.

\subsection{Setup}
\label{sec::Exp_setup}

To evaluate our proof-of-concept, we generated all mutants of the Apache Ant project (version 1.9.11\footnote{\url{https://ant.apache.org/srcdownload.cgi}}) using LittleDarwin\footnote{\url{https://github.com/aliparsai/LittleDarwin}} (version as of May 3 2018\footnote{commit id: 976ae18f6535d11bf7f66e8985fa03040c419156}). Project specific details\footnote{Gathered data and data from \url{https://www.openhub.net/p/ant}} about Ant can be found in \tabref{table:Details Ant Project}. We executed the complete test suite for each mutant and stored all of the generated reports. This allowed us to inspect which of the test cases killed which mutants. While a tool exists to automatically detect the focal methods~\cite{ghafari2015automatically, ghafari2017mining}, we will not use the tool for our proof-of-concept study, as we want our results to be independent of the tool, and to eliminate any possible errors made by the tool. As we manually needed to investigate for each test case whether the the method containing the mutant is a focal method, we only examined 423 mutants. We compare our proof-of-concept against a normal mutation execution where all test cases are considered and against a mutation execution where only all the test cases of the class from which the method that contains the mutant are considered.

We point out that we encountered some intermitted faults in the test suite. Since mutation analysis needs a passing test suite to start with, we omitted the failing test cases from our analysis. The test cases in question are located in the ``ant.AntClassLoaderTest'' class: testCodeSource, testGetPackage, testSignedJar, and in the ``ant.taskdefs.optional.XsltTest'' class: testXMLWithEntitiesInNon-AsciiPath. The error in question is a java.nio.file.InvalidPathExcept-ion: ``Malformed input or input contains un-mappable characters'' where the path includes ``\~{a}nt'' instead of ``ant''.

\begin{table}[hbt]
\centering
\caption{Details Ant Project}
\label{table:Details Ant Project}
\begin{tabular}{|l|r|}
  \hline
  Commits & 14,204	\\\hline
  Contributors & 98	\\\hline
  LOC & 229,019		\\\hline
  Test Cases & 1,777	\\\hline
  Estimated amount of effort & 59 years\footnotemark	\\\hline
  First commit & January 2000	\\\hline
  Mutant Generated & 16,354
\\\hline
\end{tabular}
\end{table}
\footnotetext{According to the COCOMO model}


\textbf{RQ1:} \textit{To what extent, using focal methods, can we identify the right mutants for a test case and vice versa?}

\noindent\textbf{Motivation.} Given the fact that we assign the responsibility to detect all possible faults in a method to a few test cases, focal methods will not detect all mutants detected by the full test suite. This can have multiple causes: a test case can be incomplete (not fulfilling its responsibilities), a test case can be missing, or the focal method approach did not establish the traceability link at method level between production code and test code.

\noindent\textbf{Approach.} We count the amount of mutants killed by the entire test suite and we count the amount of mutants killed by the reduced test suite. The difference between them is the amount of mutants the latter method missed and may be considered as false negatives. However, a deeper analysis is required here as they may also indicate poorly designed test cases suffering from the \textit{eager test} code smell~\cite{van2001refactoring}. Whether they are killed is up to the quality of the test cases in question as discussed in \secref{subsubsec:limiting}. The main goal is to assess for how many mutants we can find test cases with focal methods that contain these mutants.

\textbf{RQ2:} \textit{How much speed-up is gained by using focal methods for mutation testing?}

\noindent\textbf{Motivation.} If a mutant is detectable by a test case, then ideally that should be the only test case to be executed. If the test suite cannot detect the mutant then it is pointless to run any tests for it. Reducing the scope of the test suite ensures that when the mutant is not detectable, less resources are wasted on the mutant. Furthermore, the less test cases are considered, the earlier the test case that detects the mutant is executed and the faster the mutation testing tool becomes.

\noindent\textbf{Approach.} For any mutant, we count the amount of test cases for which the method that contains the mutant occurs as a focal method. Naturally, for the full test suite based mutation execution, all test cases are considered, and for the class based mutation execution, all test cases from that class are considered. We also count the amount of time each test case takes until the first test case that detects the mutant, this for all three mutation testing techniques.

\subsection{Results}
\label{sec::Results_Discussion}

\begin{table*}[hbt]
\centering
\caption{Results Focal Method Mutation Testing}
\label{table:Results Focal Method Mutation Testing}
\begin{tabular}{|p{2.5cm}|l|r|r|r|r|r|}
  \hline
  Class & Technique 		& Focal Mutants Detected	& False Negatives & AVG Tests Considered & Run Time & Speed-up \\\hline
  \multirow{3}{2.5cm}{ant.Ant-ClassLoader} & Full Test Suite	& 11 / 20			& 0				& 1,777.0			& 3,113.238 s & N.A. \\\cline{2-7}
  & Class Based		& 10 / 20			& 1				& 9.0			& 9.690 s & 321.3x \\\cline{2-7}
  & Focal Methods		& 9 / 20				& 2				& 1.8			& 3.082 s & 1,010.1x \\\hline\hline
  
  \multirow{3}{2.5cm}{ant.Ant-DefaultLogger} & Full Test Suite	& 4	/ 4				& 0				& 1,777.0			& 6.287 s & N.A. \\\cline{2-7}
  & Class Based		& 4	/ 4				& 0				& 1.0			& 0.010 s & 628.7x \\\cline{2-7}
  & Focal Methods		& 4	/ 4				& 0				& 1.0			& 0.010 s & 628.7x \\\hline\hline
  
  \multirow{3}{2.5cm}{ant.Ant-DirectoryScanner} & Full Test Suite	& 15	 / 17			& 0				& 1,777.0				& 785.979 s & N.A. \\\cline{2-7}
  & Class Based		& 11	 / 17			& 4				& 29	.0				& 7.221 s & 108.8x \\\cline{2-7}
  & Focal Methods		& 11	 / 17			& 4				& 24.7		& 4.486 s & 175.2x \\\hline\hline
  
  \multirow{3}{2.5cm}{ant.Ant-IntrospectionHelper} & Full Test Suite	& 14	 / 14			& 0					& 1,777.0			& 459.627 s & N.A. \\\cline{2-7}
  & Class Based		& 11	 / 14			& 3					& 14.0			& 0.433 s & 1,061.5x \\\cline{2-7}
  & Focal Methods		& 11	 / 14			& 3					& 1.1			& 0.034 s & 13,518.4x \\\hline\hline

  \multirow{3}{2.5cm}{\textbf{Total}} & \textbf{Full Test Suite}	& \textbf{44 / 55}			& \textbf{0}				& \textbf{1,777}			& \textbf{4,365.131 s} & \textbf{N.A.} \\\cline{2-7}
  & \textbf{Class Based}		& \textbf{36 / 55}			& \textbf{8}				& \textbf{15.9}			& \textbf{17.354 s} & \textbf{251.5x} \\\cline{2-7}
  & \textbf{Focal Methods}		& \textbf{35 / 55}			& \textbf{ 9}				& \textbf{8.6}			& \textbf{7.612 s} & \textbf{573.5x} \\\hline
\end{tabular}
\end{table*}

\tabref{table:Results Focal Method Mutation Testing} indicates how many mutants we found to be located in methods which we detected as focal methods. It also indicates how many of the mutants were detected by the test suite. The \textit{false negatives} indicate mutants that are not killed due to the limited amount of test cases considered by the used techniques, but that would have been killed by the full test suite. \textit{AVG tests considered} indicates how many test cases the technique can execute (on average) to detect the mutants. \textit{Run time} indicates the time needed to execute all mutants (either until the mutant is detected or all considered test cases are executed) and finally, \textit{speed-up} indicates how much faster the technique is compared to running the complete test suite.

The \textit{full test suite} technique indicates a normal mutation testing technique where the entire test suite is considered to detect the mutant. The \textit{class based} technique indicates a mutation testing technique where only the tests of the class in which the mutant is located is considered to detect the mutant. We included this to give our \textit{focal method} approach some perspective. \tabref{table:Results Focal Method Mutation Testing} contains detailed information on the gathered test classes and a global overview.

\subsubsection{\textbf{RQ1:} \textit{To what extent, using focal methods, can we identify the right mutants for a test case and vice versa?}}
In total, we examined the first 423 mutants out of 16,354. For 55 of them (13\%), we detected test cases for which the methods containing these mutants are identified as focal methods (see \tabref{table:Results Focal Method Mutation Testing}). Currently, the 368 mutants which are not found in test cases with focal methods are due to missing tests. This low recall rate is due to the way private methods are tested.
In our anecdotal experience, most developers test private methods indirectly by calling public methods which act upon them. However, the current definition of focal methods is able to identify private methods as focal methods only when such methods are executed more directly, e.g. using Java reflections\footnote{\url{https://docs.oracle.com/javase/tutorial/reflect/index.html}\label{Java reflections}}. 
In future, we plan to extend the requirements of focal methods to allow indirect private methods to become focal methods.

Of the 55 mutants located in focal methods, we see that the \textit{focal method} based approach detects fewer mutants than the \textit{class based} approach, and thus has more false negatives. The \textit{class based} and \textit{focal method} based technique detect respectively 82\% and 80\% of the mutants the full mutation testing technique detects. This is due to incomplete test cases. While some of these mutants are detected by the test suite, they are not detected by the test cases that have the responsibility to detect them. Therefore, the quality of the individual test cases in which the detected focal methods are located can be said to be 80\%.

The focal method approach benefits the most when there is an elaborate test suite where all methods are individually tested. Test suites which test at a higher level than method level, i.e. procedures and integration tests will benefit less from this technique. The extend of this impact, and the possibility to adapt focal methods to cope with procedure tests will need to be investigated in a future work.

\subsubsection{\textbf{RQ2:} \textit{How much speed-up is gained by using focal methods for mutation testing?}}
For the mutants located in focal methods, we see that the average amount of test cases considered for the mutants is drastically reduced, both for the \textit{class} based as the \textit{focal methods} based approach: 0.9\% and 0.5\%, respectively. On average, the \textit{focal methods} based approach considers half of the amount of tests the \textit{class} based approach considers. However, this highly depend on the amount of tests for the class. The more tests per class, the faster the \textit{focal methods} based approach can be compared to the \textit{class} based approach (see AntClassLoader and IntrospectionHelper in \tabref{table:Results Focal Method Mutation Testing}).

We see that the total run time of these mutants is drastically reduced as well, both for the \textit{class based} as the \textit{focal methods} based approach: respectively 0.4\% and 0.2\%. 

There are two ways to look at these speed-ups, on the one hand, we can say that this currently only impacts 13\% of the investigated mutants. Therefore, the speed-up considering all mutants is currently only 1.15x. 
On the other hand, we can say that for the methods in which 87\% of the mutants are located, the test cases that have the responsibility to detect all faults in them are missing.

The current definition of focal method requires test cases to directly call a method for it to be considered as a possible focal method. In our anecdotal experience, most developers test private methods indirectly by calling public methods which act upon them. Therefore, we plan to extend the definition of focal methods to allow indirect private methods to become focal methods as well. This will increases the amount of detected mutants and provides a drastic speed-up.

\hypobox{$\implies$ Focal methods can be used as a viable alternative to drastically reduce the test scope and run time of mutation testing, but improvements are needed to better cope with private methods.}


\section{Related Work}
\label{sec::RelatedWork}

The RIPR model (Reachability, Infection, Propagation, Revealability)~\cite{ammann2016introduction} states that in order to reveal a fault, a test case must a) reach the faulty statement (Reachability), b) cause the program state to become faulty (Infection), c) propagate the fault to the program output (Propagation) and d) cause a failure, i.e. the faulty state is asserted by the test case to its intended state (Revealability).

Three different kinds of mutation testing can be linked to this model: \textit{weak}, \textit{firm}, and \textit{strong} mutation testing. For \textit{weak} mutation testing, only the first two conditions of the RIPR model need to be satisfied. This means that a mutant is considered detected from the moment the program state of the original program and the mutated program differ. With \textit{firm} mutation testing, an extension of \textit{weak} mutation testing, the user can decide which component of the program state should differ from the original for a mutant to be considered as detected. Lastly, for \textit{strong} mutation testing, all conditions of the RIPR model need to be satisfied. This means that a mutant must influence the observable output of the program (the test oracle), and not just the program state or a component in it.

Empirical evidence~\cite{chekam2017empirical} has shown that \textit{strong} mutation testing is more powerful than weak and firm mutation testing. In essence, to detect a mutant, at least one test case of the entire test suite should fail. For faster mutation testing, one can choose to stop executing test cases as soon as a test case kills the mutant.
Ideally, we only execute those test cases for which all conditions of the RIPR model are satisfied. Test cases that do not satisfy these conditions can be excluded for performance optimisations. In the next section, we list existing techniques that consider the RIPR model to optimise for strong mutation testing.

\subsection{Existing Techniques}
Since mutation testing is such a computationally expensive technique, many researchers have sought mitigation strategies~\cite{jia2011analysis}. Many approaches originate in work on test suite minimisation, a set-cover problem that has been shown to be NP-complete -- but several approximation solutions have been proposed~\cite{usaola2010mutation}. As an example, Jeffrey and Gupta presented a test suite reduction technique with selective redundancy, a slightly more conservative approach (i.e. less reduction) that retains more of the fault detection effectiveness of the original test suite compared to previous work. Nevertheless, test suite reduction always requires a trade-off between execution time and fault detection effectiveness~\cite{shi2014balancing}.

Several regression test selection methods have been proposed to speed up mutation testing, aiming at restricting test case execution to those that target the code changes. Regression test selection methods are either dynamic (i.e. using execution information) or static (i.e. based entirely on source code analysis). Chen and Zhang performed an extensive empirical evaluation of several state-of-the-art regression test selection methods for mutation testing on 20 GitHub projects~\cite{chen2018speeding}, and conclude that the techniques are generally feasible on a file level but not for finer-grained analysis. Also, the methods studied are intended for evolving systems and not for a single version of source code.

Zhang \textit{et al.} focused on speedup of mutation testing that works for a single source code version~\cite{zhang2013faster}. They developed FaMT (Faster Mutation Testing) as an approach to prioritise and reduce the number of test cases to execute for each mutant. Inspired by research on regression test prioritisation, FaMT reorders the test cases in a way to kill the mutant earlier. Subsequently, inspired by previous work on test suite reduction, FaMT runs only the subset of test cases with a high likelihood to kill the mutant. Thus, FaMT might under-approximate the mutation score -- some of the skipped test cases might indeed have killed the mutant if they were executed.

There are also other approaches to exclude test cases from a test suite targeting a specific mutant. Bardin \textit{et al.} proposed program verification to exclude test cases that cannot reach the mutant and/or that cannot infect the program state~\cite{bardin2015sound}. Other authors have explored using (static) symbolic execution techniques to identify whether a test case can detect mutants~\cite{papadakis2012mutation, holling2016nequivack}. An example of a tool implementing this approach is PIT~\cite{coles2016pit}, that executes only those test cases that have a chance to kill the mutant, i.e. the test cases that execute the faulty statement (thus fulfilling Reachability).

\subsection{Comparison}
As seen, there already exist techniques to exclude test cases from a test suite to speed up mutation testing. The ultimate goal for these techniques, however, is to detect all the mutants that the entire test suite can detect with as few test cases as possible. In \secref{subsec:Motivation} we have argued the need for mutation testing to focus on detecting the quality of each method and its corresponding test cases instead of the overall quality of the test suite. Focal methods meet our needs, as their goal is to test the methods with the test cases that are specifically designed to test them, thus validating the quality of each test case individually.

Furthermore, focal methods deviate from the RIPR model as we will often exclude test cases that reach the faulty statement, test cases that are infected by the faulty statement, test cases that propagate the faulty statement to the program output and/or test cases that would cause a failure due to the faulty statement, but that do not have the responsibility of detecting the injected fault.

\section{Threats to Validity}
\label{sec::Threats}

As with all empirical research, we identify those factors that may jeopardise the validity of our results and the actions we took to reduce or alleviate the risk. Consistent with the guidelines for case studies research (see~\cite{case_study_research_SWE,case_study_research}), we organise them into four categories.

\noindent
\textbf{Construct validity: } did we measure what was intended?
In essence, we wanted to know whether focal methods could be used to reduce the test scope for mutation testing to drastically speed up mutation testing. Therefore, we investigated a part of the Ant project and verified that for the mutants located in methods we detected as focal methods in some test cases, there was an average speedup of 573.5x. However, we only found test cases with focal methods for 13\% (55/423) of the investigated mutants. This low recall rate is due to the fact that this leaves out most private methods, as developers mostly test these indirectly by calling public methods. In our next version, we will extend the requirements of focal methods to include indirect private methods. This will strongly influence our current results.
Using focal methods, only 35 mutants out of the 44 detectable mutants were detected. However, the 9 mutants that are detected using full mutation testing are detected in places that do not have the responsibility to detect them. These mutants are detected because they eventually altered the program state so that the fault showed up in another place. While the test cases in which these faults showed up do not necessarily have a test code smell like the \textit{eager tests} code smell, they do test methods that are not the intent of the test case. New, dedicated tests cases that have the responsibility to detect them should be written or existing ones should be expanded. It therefore does not matter that focal methods do not detect the same amount of mutants as the full mutation testing technique does, as this highlights the quality of the underlying test suite.
 
\noindent
\textbf{Internal validity:} were there unknown factors that might have affected the outcome of the analyses?
As we omitted the results from the test cases with intermittent faults, it is possible that we falsely identified some mutants as undetected while the omitted test could have detected it. Furthermore, as we manually analysed if the method where the mutant is located occurs as a focal method in some test cases, these results are subject to human error. We, however, believe that our obtained results show the viability of focal methods to reduce the test scope for mutation testing and drastically speed up mutation testing. 

\noindent
\textbf{External validity:} to what extent is it possible to generalise the findings?
The speedup of this approach comes from validating only a few test cases for each mutant (for which focal methods are found). The larger the software project and the more test cases the project has, the more beneficial this technique becomes, as the use of focal methods will always limit the considered amount of test cases to a handful. However, when the test suite has a lot of test code smells~\cite{van2001refactoring}, this might negatively impact the speedup of this approach. E.g. with the \textit{eager test} code smell, many methods are tested in a single test case. With this code smell, some methods are only tested indirectly, preventing them from becoming focal methods and thus preventing them to leverage the speedup of our proposed technique. 
Previous work that mines API usage examples from unit test cases~\cite{ghafari2017mining}, alleviated this issue by identifying focal methods within each sub-scenario of a unit test case. 

In general, this technique performs better the less test code smells the test suite has. When a project has a lot of test code smells, it might be better to first focus on removing them for better maintainability and diagnosability of the test suite~\cite{van2001refactoring}.

\noindent
\textbf{Reliability:} is the result dependent on the tools?
We explicitly chose to gather these results manually instead of using the tool to detect the focal methods~\cite{ghafari2015automatically, ghafari2017mining} to eliminate any possible inaccuracies of the tool. This allowed us to investigate the feasibility of this approach without relying on the tool to make the results reliable. The only downside is that by doing so, we limited the amount of investigated mutants, increasing the odds of skewed results. However, we believe that the results indicate that the use of focal methods is a viable approach for speeding up mutation testing.
\section{Conclusion}
\label{sec::Conclusion}

In this paper, we argued the need for mutation testing to focus on detecting the quality of each method and its corresponding test cases instead of the overall quality of the test suite. For this, a traceability link at method level between production code and test code must be established, which we achieved by using focal methods. This allows us to identify which test cases actually test which methods and vice versa.

We argued that for better diagnosability and maintainability of the test suite, it is best to test a method $f$ in a test case that is specifically designed to test $f$ and not in a test case that just so happens to call the method $f$ in one of its routines. Using focal methods, we can focus on those test cases that are designed to test specific methods and ensure their quality. This will increase the quality of each method and its corresponding test cases and not just the overall ability of the test suite to detect defects.

We demonstrated that focal methods can be used to drastically speed up mutation testing. In our limited testing of the Ant project, we observed an average speedup of 573.5x for the mutants, located in methods that we detected as focal methods in some test cases, without sacrificing accuracy. In our experiments, we noted that the test cases with the focal methods killed 80\% of the mutants killed by the entire test suite. The remaining 20\% could be detected by test cases that are not intended to test the method in which these faults where injected. New, and dedicated tests cases that have the responsibility to detect them should be written or existing ones should be expanded. It therefore does not matter that focal methods do not detect the same amount of mutants as the full mutation testing technique, as this highlights the quality of the underlying test suite -- indeed the purpose of mutation testing.

Despite these improvements, there are still opportunities for further optimisation. Currently, we did not find test cases with focal methods for 368 mutants out of the 423 investigated mutants. This low recall rate is due to the fact that the technique leaves out most private methods, as developers mostly test these indirectly by calling public methods. 
We plan to adapt the notion of focal methods such that it incorporates indirect private method calls as well. Then, a larger percentage of the executed mutants will be drastically sped up.

\section*{Acknowledgments}
\ifthenelse{\boolean{anonymous}} {
\small{This work is sponsored by the ANONYMOUS of COUNTRY under a project entitled ``ANONYMOUS".}
} {
\small{This work is supported by (a) the ITEA TESTOMATProject (number 16032), sponsored by VINNOVA --- Sweden's innovation agency; (b) Flanders Make vzw, the strategic research centre for the manufacturing industry.

We also gratefully acknowledge the financial support of the Swiss National Science Foundation for the project ``Agile Software Analysis'' (SNSF project No.\,200020-162352, Jan 1, 2016 - Dec.\,30, 2018).
}
}

\newpage 

\ifthenelse{\boolean{acmtemplate}} {
	\bibliographystyle{ACM-Reference-Format}
}{
	\bibliographystyle{IEEEtran}
}
\bibliography{mybib}

\end{document}